\pdfoutput=1
\documentclass[a4paper,american,floatfix,pdftex,superscriptaddress,twoside,%
aps,prx,%
longbibliography,
reprint,final
]{revtex4-1}%
\usepackage{amsfonts,amsmath,amssymb}
\usepackage{commath}
\usepackage[T1]{fontenc}
\usepackage{graphicx}%
\usepackage[utf8]{inputenc}
\usepackage{mathtools}
\usepackage{microtype}
\usepackage[obeyFinal,textsize=footnotesize]{todonotes}
\usepackage{xspace}
\usepackage{hyperref,hypernat}
\usepackage[todos]{CMImacros}

\usepackage{acronym}
\newacro{TDSE}{time-dependent Schrödinger equation}
\newacro{SFA}{strong-field approximation}
\newacro{HOMO}{highest occupied molecular orbital}
\newacro{FWHM}{full width at half maximum}
\newacro{ATI}{above-threshold ionization}
\newacro{ADK}{Ammosov-Delone-Kra\v{\i}nov}
\newacro{VMI}{velocity-map imaging}

\graphicspath{{./figures/}} 
\hypersetup{citebordercolor=yellow,linkbordercolor=red,urlbordercolor=blue,hidelinks}

\newcommand{\cfeldesy}{\affiliation{Center for Free-Electron Laser Science, Deutsches
      Elektronen-Synchrotron DESY, Notkestraße 85, 22607 Hamburg, Germany}}%
\newcommand{\uhhchem}{\affiliation{Department of Chemistry, Universität Hamburg,
      Martin-Luther-King-Platz 6, 20146 Hamburg, Germany}}%
\newcommand{\uhhcui}{\affiliation{Center for Ultrafast Imaging, Universität Hamburg, Luruper
      Chaussee 149, 22761 Hamburg, Germany}}%
\newcommand{\uhhphys}{\affiliation{Department of Physics, Universität Hamburg, Luruper Chaussee 149,
      22761 Hamburg, Germany}}%
\newcommand{\jkemail}{\email[Email: ]{jochen.kuepper@cfel.de}}%
\newcommand{\cmiweb}{\homepage[URL: ]{https://www.controlled-molecule-imaging.org}}%

\DeclareMathOperator{\tr}{tr}

\begin{document}
\title{Strong-field ionization of complex molecules}%
\author{Joss Wiese}\cfeldesy\uhhcui\uhhchem%
\author{Jolijn Onvlee}\cfeldesy\uhhcui%
\author{Sebastian Trippel}\cfeldesy\uhhcui%
\author{Jochen Küpper}\jkemail\cmiweb\cfeldesy\uhhcui\uhhchem\uhhphys
\begin{abstract}\noindent%
   Strong-field photoelectron momentum imaging of the prototypical biomolecule indole was
   disentangled in a combined experimental and computational approach. Experimentally, strong
   control over the molecules enabled the acquisition of photoelectron momentum distributions in the
   molecular frame for a well-defined, narrow range of incident intensities. A novel, highly
   efficient semiclassical simulation setup based on the adiabatic tunneling theory quantitatively
   reproduced these results. Jointly, experiment and computations revealed holographic structures in
   the asymptotic momentum distributions, which were found to sensitively depend on the alignment of
   the molecular frame. We identified the essential molecular properties that shape the
   photoelectron wavepacket in the first step of the ionization process and employ a
   quantum-chemically exact description of the cation during the subsequent continuum dynamics. The
   detailed modeling of the molecular ion, which accounts for its polarization by the laser-electric
   field, enables the simulation of laser-induced electron diffraction off large and complex
   molecules and provides full insight into the photoelectron's dynamics in terms of semiclassical
   trajectories. This provides the computational means to unravel strong-field diffractive imaging
   of biomolecular systems on femtosecond time scales.
\end{abstract}
\maketitle


\section{Introduction}
\label{sec:introduction}
Strong-field ionization is a versatile and powerful tool for the imaging of molecular structure,
allowing simultaneous access to valence-electronic topology and atomic positions with
sub-atomic-unit resolution in time and space~\cite{Corkum:NatPhys3:381}. A generalization of the
underlying imaging techniques, which were born in the field of atomic physics, to complex
biomolecular targets promises deep insights into the biochemical machinery of life and this
extension is the aim of the current work.

In strong-field-ionization experiments the desired structural information can be obtained in two
complementary ways: through the observation of the photoelectron itself, which is discussed in
detail below, or through the burst of light that is emitted when the electron recombines with the
cation, referred to as high-harmonic generation. The latter was employed in many studies on atoms
and molecules unraveling valence-orbital~\cite{Itatani:Nature432:867} and atomic
structure~\cite{Kanai:Nature435:470}, but is not discussed further in our work.

The current article focuses on the photoelectron wavepacket as it picks up molecular structural
imprints during its formation and motion in the continuum. The entire strong-field process may
descriptively be resolved into three stages, which are linked through the photoelectron's
propagation in the combined time- and position-dependent field that is exerted by the laser and the
cation:

(i) The photoelectron is born in the continuum. Its wavefunction is shaped by the ionization
potential~\cite{Ammosov:SVJETP64:1191} and the nodal geometry~\cite{Lein:JPB36:L155,
   Spanner:JPB37:L243, Holmegaard:NatPhys6:428, Trabattoni:cutoff:inprep} of the initial bound
electronic state.

(ii) Upon propagation within close vicinity to the ionized target the photoelectron wavepacket picks
up additional information about the cation through interaction with its electric potential. For
circularly polarized laser pulses this encounter is essentially limited to early times right after
birth in the continuum and, if the target molecule is chiral, results in photoelectron circular
dichroism~\cite{Lux:ACIE51:5001, Dreissigacker:PRA89:053406, Janssen:PCCP16:856}. For linear
polarization, however, the photoelectron is, in first instance, quickly driven away from the
molecule yet might partly return at a later time. If the momentum at return is rather small, \ie,
comparable to the initial momentum distribution at birth, the wavepacket may holographically
interfere with another portion of itself that does not return to the cation. Those holographic
interferences encode the tunnel exit position and the initial momentum distribution of the
photoelectron~\cite{Huismans:Science331:61, Meckel:NatPhys10:594, Walt:NatComm8:15651}. However, if
the re-collision occurs at the maximum of the kinetic energy, namely about three times the
ponderomotive potential~\cite{Paulus:JPB27:L703}, the photoelectron may intrude deeply into the
ionic field probing the instantaneous positions of the nuclei. This process includes only a
temporally narrow slice of the initial wavepacket that returns after less than one optical cycle
near a zero-crossing of the laser field. It is referred to as laser-induced electron diffraction
\cite{Zuo:CPL159:313, Blaga:Nature483:194, Pullen:NatComm6:7262, Amini:PNAS116:8173,
   Karamatskos:JCP150:244301}.

(iii) After both, the decay of the laser field and the escape from the cation's Coulomb potential,
are completed, the photoelectron wavepacket assumes its asymptotic form in momentum space. This is
the stage one can readily access experimentally. It is highly desirable to extract unambiguous
information about the target molecule, that was imprinted onto the photoelectron, from this
detectable final momentum distribution. Although attempts of inversion recipes were formulated
already, building for example upon the quantitative rescattering theory~\cite{Chen:PRA79:033409}
for the evaluation of laser-induced electron diffraction experiments, they rely on severe
approximations regarding the initial photoelectron wavepacket or its continuum motion. For
instance, the returning electron's momentum distribution is frequently assumed to be free of any
angular modulation. But since molecular systems may give rise to rather structured initial electron
waves~\cite{Holmegaard:NatPhys6:428, Dimitrovski:JPB48:245601, Paul:PRL120:233202,
   Schell:SciAdv4:eaap8148}, the target-dependent formation and continuum motion of the
photoelectron needs to be well understood in order to come up with a general inversion technique.

During the last twenty years numerous models have been brought into being, which elegantly tackle
various different aspects of strong-field physics. The numerical evaluation of the
\ac{TDSE}~\cite{Bauer:ComPhysComm174:396, Bian:PRA84:043420} represents the most rigorous
approach among them, but remains mostly restricted to atomic systems. Further approaches to tackle
strong-field ionization of complex molecules rely on the time-dependent density-functional
theory~\cite{Wopperer:EPJB90:2017, Trabattoni:cutoff:inprep}. They can be efficiently applied for
wavelengths up to the mid-infrared regime, but so far do not allow to track the photoelectron's
dynamics in the continuum.

A broad class of models building upon the \ac{SFA} in the single-active electron picture allows for
the treatment of more complex targets. Those \acs{SFA}-based theories commonly divide the whole
process into two steps corresponding to stage (i) and stages (ii+iii) described above. In the
low-frequency limit, the initial continuum wavefunction can be described by means of quasistatic
tunneling theories~\cite{Ammosov:SVJETP64:1191}, which can be step-wise supplemented to grasp
non-adiabatic behavior during tunneling~\cite{Li:PRA93:013402} and allow for the consideration of
imprints from the initial bound electronic state~\cite{Spanner:JPB37:L243,
Holmegaard:NatPhys6:428}. An alternative, wavelength-independent description of the early
photoelectron wavefunction can be obtained through the sub-barrier Coulomb-corrected
\acs{SFA}~\cite{Yan:PRA86:053403}.

The very classical nature of the electron's motion in continuum has stimulated several studies,
which rely on its decomposition into many classical trajectories. A careful analysis of these
trajectories can then help to gain an illustrative understanding of intricate strong-field
phenomena~\cite{Shvetsov-Shilovski:PRA97:013411}. The two most important approaches to describe the
electron's phase in the continuum are the frequently used Coulomb-corrected
\acs{SFA}~\cite{Huismans:Science331:61, Yan:PRA86:053403} and semiclassical time-dependent
propagators~\cite{Shvetsov-Shilovski:PRA94:013415}.

We note that in the vast majority of published models the cation is approximated by a point charge
during the continuum propagation of the freed electron.

Here, we introduce a novel and very efficient semiclassical simulation setup. Our approach builds
upon the adiabatic tunneling theory and allows for the consideration of structural imprints from the
initially bound orbital. The continuum propagation of the photoelectron wavepacket is treated
through semiclassical trajectories that experience the detailed, quantum-chemically exact, field of
the cation.

Due to its efficiency our model is ideally equipped for disentangling the structures of even highly
complex molecules through strong-field ionization. In conjunction with a strong-control
experimental scheme, which enabled a well-defined ionization study in the molecular frame, we
present a combined experimental and computational survey on the strong-field ionization of the
prototypical biomolecular building block indole (C$_8$H$_7$N), which is the major
ultraviolet-absorbing chromophore of proteins.

Unless stated otherwise, atomic units (at.~u.) are used throughout this article.

\section{Semiclassical model}
\label{sec:model}
\begin{figure}
   \includegraphics[width=\linewidth]{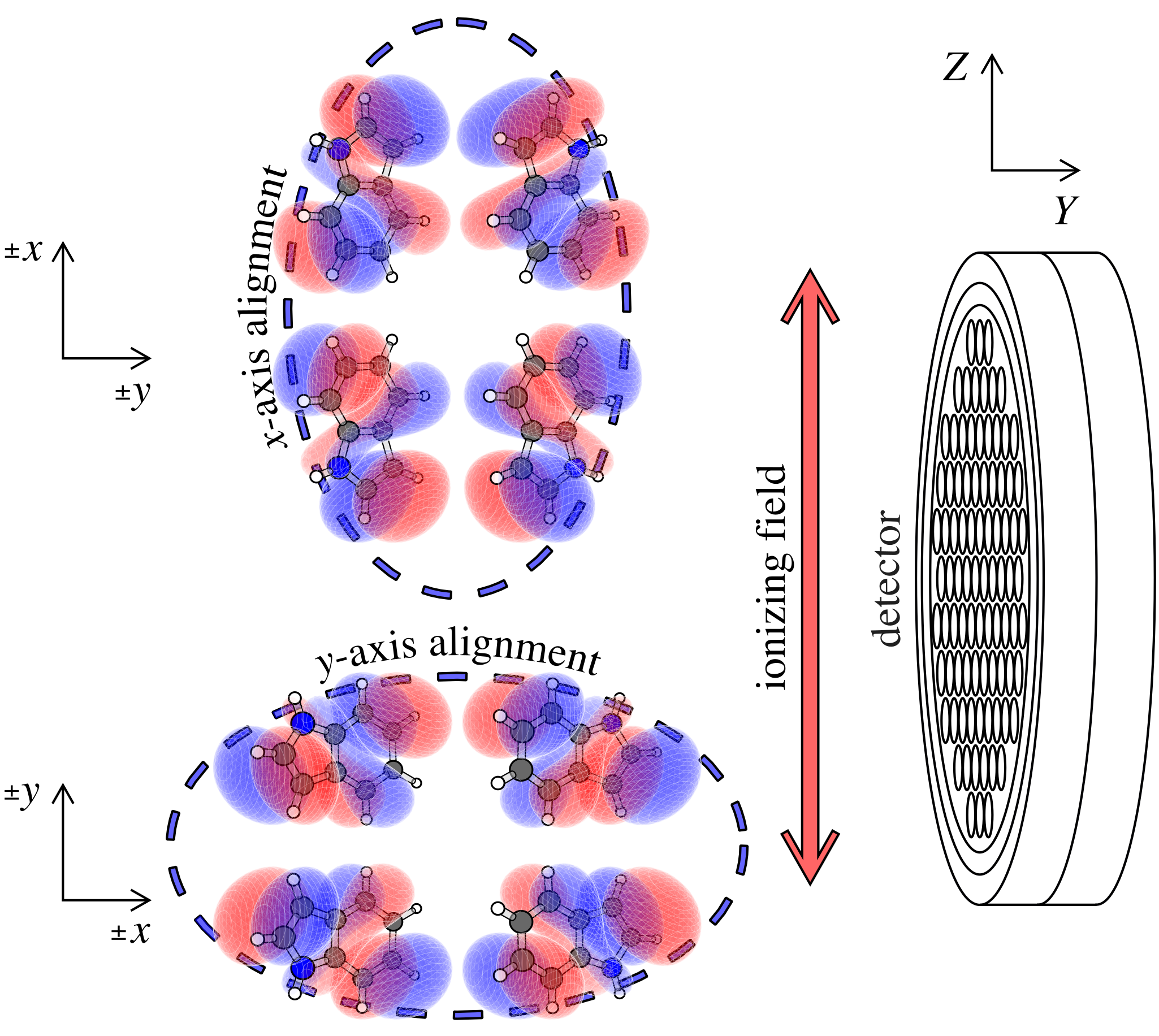}
   \caption{Indole molecules were three-dimensionally fixed in the laboratory frame according to two
      different alignments of their principal-polarizability frame, depicted on the left, with
      respect to the polarization axis of the ionizing laser field, depicted as double-headed red
      arrow. In the experiment, this was achieved through irradiation with a nonresonant laser field
      carrying elliptical polarization, sketched by the blue ellipses. Both laser beams propagated
      collinearly along the $X$ axis. The molecule's \acs{HOMO} is shown in an isosurface
      representation for all four degenerate orientations.}
   \label{fig:alignment}
\end{figure}
Our description of the strong-field ionization process employs a two-step approach: Treating the
photoelectron wavefunction at birth in the continuum through quasistatic tunneling theory allows
for the analytical description of the first step, which is explained in \autoref{sec:psi0}.
Subsequently, the final wavefunction is found by sampling its dynamics in the combined electric
field of the laser and the polarizable cation on plane-wave quantum orbits, which is depicted in
\autoref{sec:contprop}, while \autoref{sec:psiinf} illustrates the procedure to compose the 
probability density of the final wavefunction from the asymptotic trajectory properties.

We tested our model by simulating the photoelectron momentum distribution from metastable xenon at a
wavelength of $7~\um$, which yielded excellent agreement with the published experimental and
\ac{TDSE} results~\cite{Huismans:Science331:61}.

Throughout this manuscript the reference coordinate frame will be set through the polarization of
the laser field, \ie, laboratory coordinate frame, with the $Z$ axis corresponding to the
polarization axis. The molecular frame is assumed to be linked to the laboratory frame \emph{via}
the indole molecule's principal axes of polarizability, $\alpha_{xx}>\alpha_{yy}>\alpha_{zz}$. Two
different molecular alignment configurations will be used, $x$-axis alignment with
$(z,\pm{y},\pm{x})\rightarrow(X,Y,Z)$ and $y$-axis alignment with
$(z,\pm{x},\pm{y})\rightarrow(X,Y,Z)$. Note, that the signs of $x$ and $y$ axes are not fixed,
giving rise to four simultaneous orientations, while the sign of the $z$~coordinate is
insignificant as a result of the molecule's $C_\text{s}$ point group. See \autoref{fig:alignment}
for a scheme of all four orientations for both alignment cases in the laboratory coordinate frame.

\subsection{Initial photoelectron wavefunction}
\label{sec:psi0}
The quasistatic \ac{ADK} tunneling theory~\cite{Ammosov:SVJETP64:1191}, empirically extended to the
barrier-suppression regime~\cite{Tong:JPB38:2593}, yields the instantaneous ionization rate $w$. In
conjunction with the initial momentum distribution in the plane orthogonal to the polarization
axis, the probability density of the initial photoelectron wavefunction~\cite{Liu:PRL116:163004}
can be obtained:
\begin{equation}
   \left| \psi_0 \right|^2 \propto w(t) \cdot \eta^2(t,p_X) \cdot \xi(t) \cdot
   \text{e}^{-\xi(t) \cdot \left( p_X^2 + p_Y^2 \right)} \,. \label{eq:psi0}
\end{equation}
$\xi=\sqrt{2\Ip{}}/\epsilon$ is the ratio of the initial state's characteristic momentum, \ie, the
square root of twice the ionization potential \Ip{}, and the instantaneous field strength
$\epsilon$. $\eta$ represents the electronic-structure imprint of the initial bound state onto the
momentum distribution of $\psi_0$ and can be acquired by linking the momentum in the initial
state $\vec{k}$ to the momentum at the tunnel exit $\vec{p}$ \emph{via} the temporal saddle point
equation~\cite{Liu:PRL116:163004}:
\begin{align}
  \begin{split}
     k_X &= p_X \\
     k_Y &= p_Y \\
     k_Z &= p_Z + A_Z(t_s) = \pm i \sqrt{2\Ip{} + p_X^2 + p_Y^2} \,.
  \end{split}
\end{align}
Here, $A_Z(t_s)$ is the $Z$~component of the vector potential at the complex saddle-point time
$t_s$. Subsequently, the corresponding molecular electronic structure factor can be reorganized to
the form $\eta \cdot \text{e}^{i\Delta\phi_0}$, with $\eta$ providing a momentum imprint onto
$\psi_0$ and $\Delta\phi_0$ introducing an initial phase offset.

A simplified description of the \ac{HOMO} as a p$_X$~orbital,
\begin{align}
  \begin{split}
     \psi_0 &\propto \frac{k_X}{\sqrt{k_X^2 + k_Y^2 + k_Z^2}}\\
     &\propto \frac{p_X}{\sqrt{2\Ip{}}} \cdot \text{e}^{-i\pi/2} \,,
  \end{split}
\end{align}
results in a constant phase shift and a momentum imprint
\begin{equation}
   \eta(t,p_X) = \frac{p_X}{\sqrt{2\Ip{}(t)}} \,.
\end{equation}

Within the scope of this model description it is sufficient to draw tuples of
$\left(t_0,p_{0X},p_{0Y}\right)$ from $|\psi_0|^2$ to get complete sets of initial phase-space
coordinates, since the tunnel exit $\vec{r}_0$ and $p_{0Z}$ can be unambiguously inferred.

$\vec{r}_0$ is deduced from the field-direction model~\cite{Pfeiffer:NatPhys8:76} and the
corresponding momentum component along the polarization axis, $p_{0Z}$, can be accessed with the aid
of the adiabatic tunneling theory embodying the first nonadiabatic
correction~\cite{Li:PRA93:013402}. Within the barrier-suppression regime the immanent excess energy
is transformed into additional longitudinal momentum~\cite{Pfeiffer:NatPhys8:76}.

Finally, starting from the instant of birth, the plane-wave phases along the quantum orbits, which
are employed to trace the wavefunction's dynamics, are given
by~\cite{Shvetsov-Shilovski:PRA94:013415}
\begin{equation}
   \phi(t_1) = -\vec{p}_0 \!\cdot\! \vec{r}_0 + \Ip{} \!\cdot\! t_0 -\int_{t_0}^{t_1} \!\!
   \left( \frac{p^2(t)}{2} - \frac{2}{r(t)} \right) \text{d}t \label{eq:phi} \,.
\end{equation}
The second term of the integrand in the action integral of~\eqref{eq:phi},
$-2/r=V(\vec{r})-\vec{r}\cdot\vec{\nabla}V(\vec{r})$, originates from a pure far-field description
of the potential energy, $V=-1/r$. In order to also properly track the phases of those quantum
trajectories that intrude deeply into the cationic potential, one would need to model both,
$V(\vec{r})$ and $\vec{\nabla}V(\vec{r})$, in quantum-chemical detail. However, due to the large
de~Broglie wavelength of the photoelectron wavepacket under the present experimental conditions,
\emph{vide infra}, we approximate $V(\vec{r})$ by this coarse description without losing agreement
with the experimental data.

\subsection{Continuum propagation}
\label{sec:contprop}
In order to acquire the wavefunction's asymptotic probability density in momentum space, the
classical equations of motion are numerically solved within the total combined electric field of the
laser and the cation. For the sake of numerical efficiency, the ionic field
$-\vec{\nabla}V(t,\vec{r})$ is computed at varying levels of accuracy, in three different spatial
domains: At large distances from the center of nuclear charge a multipole description of the
electric field, including monopole, permanent dipole moment, and induced dipole moment, is employed,
in close proximity to the nuclei the total electric field is approximated by an electric monopole
with an atom-dependent effective charge, fully neglecting the field of the laser, and elsewhere the
quantum-chemically exact electric field is utilized. See \autoref{sec:numsetup} for the spatial
boundaries of these domains.

In the close-proximity regime the application of Kepler's laws of orbital
mechanics~\cite{Landau:Mechanics} to the resulting effective two-body problem allows for an
analytical description of the orbital motion and provides the means to sort out trajectories that
end up on stationary orbits around the ion. If the total energy of such a two-body system is
negative, recombination of electron and cation is assumed and the corresponding quantum orbit does
not contribute to the asymptotic photoelectron wavefunction.

Starting from a quantum orbit's individual time of birth in the continuum, $t_0$, the ordinary
differential equations for position, momentum, and phase
\begin{align}
  \begin{split}
     \text{d}\vec{r} &= \vec{p} \,\text{d}t \\
     \text{d}\vec{p} &= -\left( \vec{\epsilon}(t) - \vec{\nabla}V(t,\vec{r}) \right) \text{d}t \\
     \text{d}\phi &= -\left( \frac{p^2}{2} - \frac{2}{r} \right) \text{d}t
     \label{eq:odes}
  \end{split}
\end{align}
are numerically integrated until the time when the laser field is fully decayed. The asymptotic
momentum and the relative phase can then be obtained through Kepler's
laws~\cite{Shvetsov-Shilovski:PRA94:013415}.

\subsection{Asymptotic wavefunction}
\label{sec:psiinf}
From the asymptotic momenta and relative phases of all available quantum orbits the probability
density of the final photoelectron wavefunction in momentum space, $|\psi_\infty(\vec{p})|^2$, is
then composed as follows. First, the momentum-space volume of interest is divided into bins in
spherical energy coordinates:
\begin{align}
  \begin{split}
     U &= \frac{1}{2}\left( p_X^2 + p_Y^2 + p_Z^2 \right)\\
     \text{cos}(\theta) &= \frac{p_Y}{\sqrt{2U}}\\
     \text{tan}(\varphi) &= \frac{p_X}{p_Z} \,.
  \end{split}
\end{align}
Subsequently, for each bin $\left( U_j,\theta_k,\varphi_l \right)$ the modulus of the coherent sum
over all associated $M_{jkl}$ asymptotic plane waves is computed:
\begin{equation}
   \left| \psi_\infty(U_j,\theta_k,\varphi_l) \right|^2 = \left| \sum_{m=1}^{M_{jkl}}
      \text{e}^{i\phi_{jklm}} \right| \label{eq:cohsum}.
\end{equation}
Before generating the modulus of the bin-wise coherent sum, the complex-phase histogram may be
coherently symmetrized to exploit the intrinsic symmetry of the photoelectron wavefunction; see
\autoref{sec:numsetup} for the detailed treatment of the three-dimensionally aligned indole
molecule. After interpolation on a Cartesian momentum-space grid the asymptotic momentum map is then
obtained through summation along the $Y$ axis, which mimics the momentum projection in a \ac{VMI}
spectrometer.

The use of spherical energy coordinates comes with three major advantages: (i) The curvatures of the
bin edges resemble the shapes of the interference structures of $|\psi_\infty|^2$. (ii) The volume
element scales with $p\cdot\text{sin}(\theta)$, giving rise to larger bins with increasing radial
momentum, which partly makes up for the smaller number of trajectories. (iii) The inherent energy
coordinate allows for a straightforward emulation of focal-volume averaging.

While (i) and (ii) result in a highly increased convergence of $\abs{\psi_\infty}^2$ as a function
of the number of samples, (iii) enables the efficient handling of a distribution of incident
intensities as it is often encountered in strong-field experiments. Within a narrow range of
incident intensities and in absence of electronic resonances the change of $|\psi_\infty|^2$ with
respect to intensity is dominated by the ponderomotive shift of the nonresonant \ac{ATI}
interferences~\cite{Freeman:JPB24:325}. Those \ac{ATI} structures represent a purely radial pattern
with an energy spacing equal to the photon energy, $\hbar\omega$, between adjacent maxima. Thus, a
small uncertainty in the ponderomotive energy due to focal-volume averaging can easily be emulated 
by introducing this uncertainty to the energy axis of $|\psi_\infty|^2$, which smears out the 
\ac{ATI} pattern but leaves the angular structures unchanged.

\section{Computational setup}
\label{sec:numsetup}
\begin{table}
   \setlength{\tabcolsep}{5mm}
   \centering
   \begin{tabular}{lrc}
     \hline
     orbital & \multicolumn{1}{c}{\centering energy (eV)} & $\chi$ \\
     \hline
     HOMO &  $-7.83$ & A$''$\\
     HOMO$-1$ &  $-8.24$ & A$''$\\
     HOMO$-2$ & $-10.52$ & A$''$\\
     HOMO$-3$ & $-12.78$ & A$''$\\
     \hline
   \end{tabular}
   \caption{Field-free energies and irreducible representations of the four highest occupied
      molecular orbitals of indole. Energies were calculated at the MP2/aug-cc-pVTZ level of theory
      using \texttt{GAMESS}~\cite{Schmidt:JCC14:1347, Gordon:GAMESS:2005}.}
   \label{tab:emo}
\end{table}
\begin{figure}
   \includegraphics[width=\linewidth]{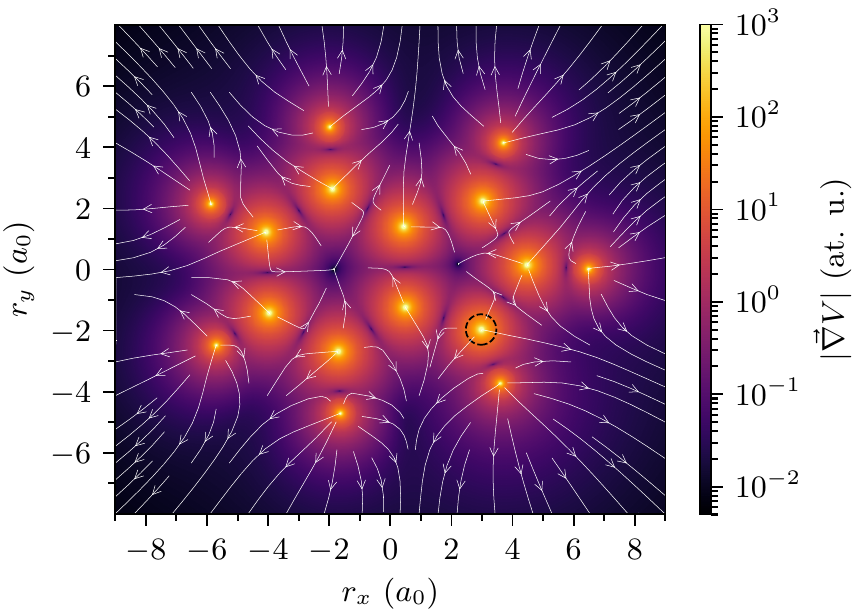}
   \caption{The quasistatic electric field exerted by the indole cation is shown for an external
      laser field $\vec{\epsilon} = 0$ for the slice in the molecular plane. Positions are specified
      with respect to the neutral molecule's principal axes of polarizability. The field strength is
      mapped onto a logarithmic color scale and field lines are depicted in white. The dashed black
      circle around the N atom illustrates the zone of influence for the two-body approximation,
      which is employed within a radius of \bohr{0.5} around the nuclei. For comparison, the peak
      electric field of the ionizing laser was $\epsilon_0=2.9\times10^{-2}$.}
   \label{fig:qef}
\end{figure}
Ionization of the indole molecule was described as occurring from a single p$_X$~orbital, which
resembles \ac{HOMO} and \acs{HOMO}$-1$ both in energy and symmetry; see \autoref{tab:emo}. Already
for \acs{HOMO}$-2$, 2.7~eV below the \ac{HOMO}, the corresponding \ac{ADK} tunneling rate is
roughly three orders of magnitude smaller. So contributions from all lower-lying orbitals were
readily neglected.

The electric field of the ionizing laser was defined as
\begin{equation}
   \epsilon_Z(t) =
   \begin{cases*}
      \epsilon_0 \cdot \cos^2\!\left(\frac{\omega t}{2n}\right) \cdot \cos(\omega t) & if
      $|\omega t| < n\pi$ \\
      0 & otherwise
   \end{cases*}
\end{equation}
with a cosine-square envelope that for $n=44$ reproduced the \ac{FWHM} of the intensity envelope of
the experimental pulses. A peak electric field of $\epsilon_0=2.9\times10^{-2}$ was utilized,
corresponding to the larger of the two pulse energies employed in the experiment, \emph{vide infra}.

The ionization potential of the molecule was computed considering Stark shifts up to second
order:
\begin{equation}
   \Ip{}(\vec{\epsilon}) = \Ip{}^{(0)} - \Delta \vec{\mu} \!\cdot\! \vec{\epsilon} - \frac{1}{2}
   \vec{\epsilon}^\text{ T}\! \Delta \alpha \, \vec{\epsilon} \, ,
\end{equation}
with the field-free ionization potential $\Ip{}^{(0)}$ and the differences of dipole moment
$\Delta\vec{\mu}$ and static polarizability $\Delta\alpha$ between cationic and neutral species.
These properties were computed using the \texttt{GAMESS} quantum chemistry
software~\cite{Schmidt:JCC14:1347, Gordon:GAMESS:2005} at the MP2/aug-cc-pVTZ level of theory, see
\autoref{sec:qcres} for details. Throughout the numerical treatment, the cation was frozen in the
nuclear equilibrium geometry of the neutral molecule.

Samples of initial phase-space coordinates were generated through rejection
sampling~\cite{Press:NumRecF77:1992} from the probability density \eqref{eq:psi0}. The tunnel-exit
positions and possible additional momentum offsets along the polarization axis in the
barrier-suppression regime were obtained based on quantum-chemistry calculations using
\texttt{Psi4}~\cite{Parrish:JCTC13:3185} at the HF/aug-cc-pVTZ level of theory, following the
field-direction model, see \autoref{sec:fielddir} for details. For the laser intensity utilized, the
ionization process was found to enter the barrier-suppression regime close to the peak electric
field for the $x$-axis alignment case. In contrast, for $y$-axis alignment the tunnel exit tightly
follows the simple description
$\vec{r}_0=-\vec{\epsilon}\,\Ip{}/\epsilon^2$~\cite{Popruzhenko:JPB47:204001}, which considers the
laser field to be quasistatic and ignores the Coulomb distortion of the barrier.

Within a $(\bohr{60})^3$ cube, centered at the molecule's center of nuclear charge, \emph{ab
initio} quantum chemistry calculations were used to describe the electric field of the cation, see
\autoref{sec:catfield}. This detailed modeling of the ion's field is necessary, because the indole
cation experiences strong polarization in response to the laser-electric field. Such behavior is
expected to be general for molecules with $\pi$-conjugated electrons and, thus, makes this
external-field dependent description indispensable for biomolecular systems. For both, the
application of the field-direction model and the evaluation of the quasistatic field of the ion,
quantum-chemical calculations were conducted beforehand for discrete positions and external fields,
see \autoref{sec:catfield} for details. Further values were obtained through interpolation in the
semiclassical propagation setup.

At distances $\bohr{<\!0.5}$ with respect to any of the molecule's nuclei the total electric field
was modeled using an electric monopole with an atom-dependent effective charge. For the H, C, and N
atoms in indole effective charges of $+1$, $+4$ and $+5$ were used. These values were retrieved by
finding the integer-charged monopoles that fitted the quantum-chemically exact field best on
spherical surfaces with $\bohr{0.5}$ radius around the atomic centers. \autoref{fig:qef} shows a
slice through the total electric field that is exerted by the indole cation.

The equations of motion~\eqref{eq:odes} were integrated following an embedded Runge-Kutta scheme
with adaptive step size (Cash-Karp)~\cite{Press:CompPhys6:188}. For each alignment case roughly
$5\times10^9$ non-recombining quantum orbits were traced. Due to the use of the spherical energy
coordinate system introduced in \autoref{sec:psiinf} roughly $5\times10^8$ plane-wave samples
sufficiently reconstructed all relevant radial and angular features of $|\psi_\infty|^2$.

To account for the coherent superposition of all four degenerate orientations in $\psi_\infty$ the
symmetry of $\psi_\infty$ with respect to inversion of the $X$ and $Y$ axes was exploited \emph{a
   posteriori} by coherently symmetrizing the histogram of bin-wise coherent sums in
\eqref{eq:cohsum}. The $X$-axis symmetry is a result of the molecule's $C_\text{s}$ point group and
the $Y$-axis inversion transforms two orientations into each other. The $Z$-axis symmetry of the
aligned molecular ensemble could not be exploited in the same way, as the corresponding axis
inversion would also flip the sign of the laser-electric field. As a consequence, for each of the
two alignment scenarios two orientation cases were independently simulated and coherently added.

\section{Experimental setup}
\label{sec:expsetup}
The experimental setup was described in detail elsewhere~\cite{Trippel:MP111:1738, Chang:IRPC34:557,
   Trabattoni:cutoff:inprep}. In brief, a cold molecular beam was created by supersonically
expanding indole seeded in 95~bar of helium through an Even-Lavie valve~\cite{Even:EPJTI:2:17}
operated at a repetition rate of 100~Hz and heated to \celsius{110}. Using an electrostatic
deflector an ultracold high-purity sample was obtained~\cite{Filsinger:JCP131:064309}. Employing
elliptically polarized laser pulses at a peak intensity of $1\times10^{12}~\Wpcmcm$, which carried
a saw-tooth temporal shape with a rise time of 500~ps, the molecular ensembles were aligned within
the center of a \ac{VMI} spectrometer~\cite{Eppink:RSI68:3477}. These alignment laser pulses, which
were spectrally centered at 800~nm and had a polarization ellipticity of 3:1 in intensity, allowed
for the nonresonant, quasi-adiabatic fixation~\cite{Trippel:MP111:1738, Trippel:PRA89:051401R} of
the indole molecule's principal axes of polarizability in the laboratory frame. Two different
alignment scenarios, \autoref{fig:alignment}, could be realized by rotating the polarization with a
half-wave plate. The resulting degree of alignment was estimated to be
$\expectation{\cos^2\delta}\approx0.9$~\cite{Mullins:indole:inprep}.

A second laser pulse, spectrally centered around 1300~nm, with a duration of 70~fs (\acs{FWHM}) and
linear polarization (ellipticity 200:1), singly ionized the molecules and photoelectron momentum
maps were recorded using a high-energy \ac{VMI} spectrometer. A position-sensitive detector,
consisting of a microchannel-plate stack, a phosphor screen, and a high-frame-rate camera, was used
for counting and two-dimensional momentum-mapping of individual electrons. To lower the impact of
focal-volume averaging onto the incident-intensity distribution, intensity-difference
photoelectron-momentum maps~\cite{Wiese:NJP21:083011} were obtained using peak intensities of $2.5$
and $3.0\times10^{13}~\Wpcmcm$. The resulting range of the ponderomotive potential
$\Delta\Up{}=0.83~\hbar\omega=0.79$~eV between these two peak intensities is still too large to
resolve nonresonant \ac{ATI} structures but represents a compromise between incident-intensity
resolution and the amount of signal that is left in the intensity-difference momentum maps. The
maximum kinetic energy at which the photoelectrons could re-encounter the
cation~\cite{Paulus:JPB27:L703}, $\Up{3.17}\approx15$~eV, resulted in a minimum de~Broglie
wavelength of \bohr{\ordsim6}, justifying the coarse description of the ion's potential energy in
\autoref{sec:psi0}.

\section{Results and Discussion}
\label{sec:results}
\begin{figure}
   \includegraphics[width=\linewidth]{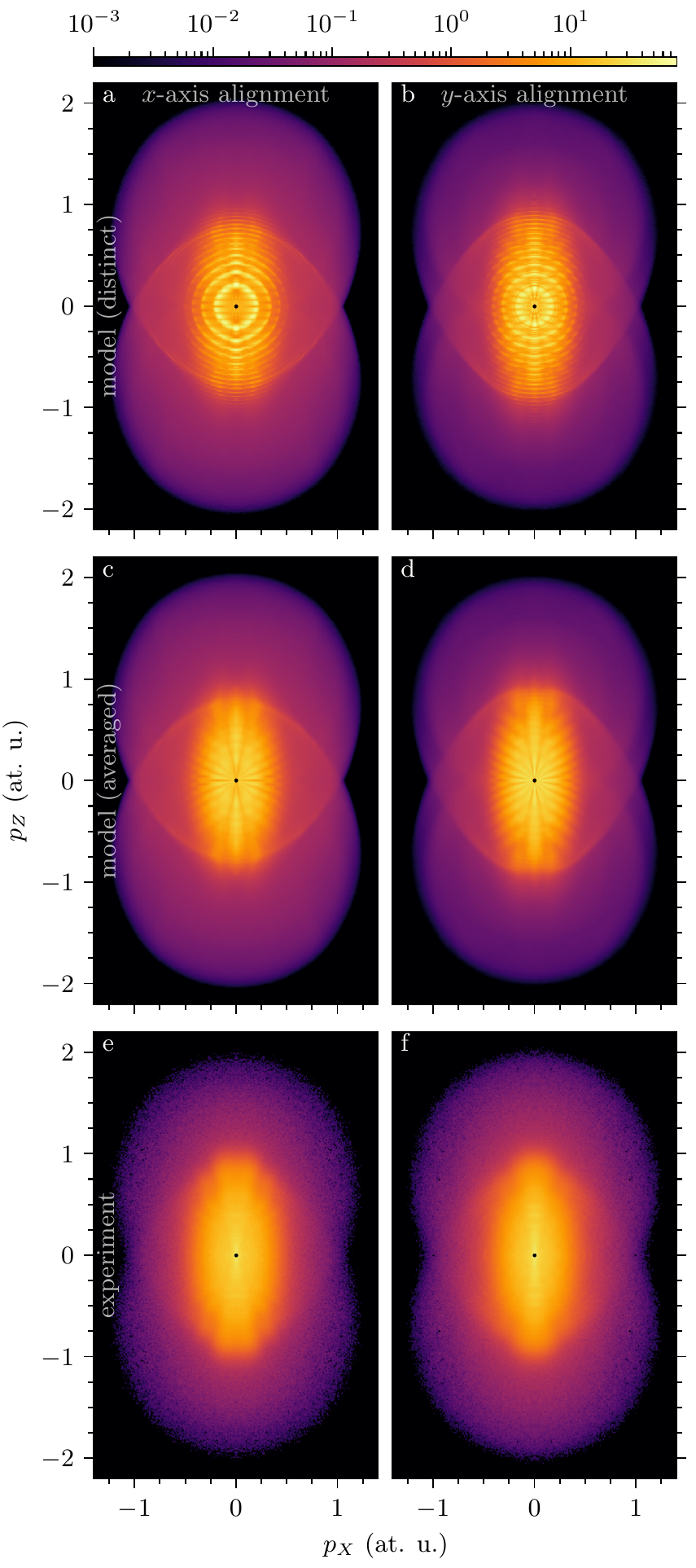}
   \caption{Asymptotic photoelectron-momentum maps for the $x$-axis alignment case in the left and
      the $y$-axis alignment case in the right column. The top row shows momentum maps modeled for a
      distinct incident intensity and the middle row shows the corresponding maps with focal-volume
      intensity averaging; see text for details. The bottom row shows the experimental results. All
      images were normalized to their mean intensity and are displayed on a common logarithmic color
      scale shown at the top. See \autoref{fig:pmaps_zoom} for a zoom into the central part of the
      experimental data.}
   \label{fig:pmaps}
\end{figure}
\begin{figure}
   \includegraphics[width=\linewidth]{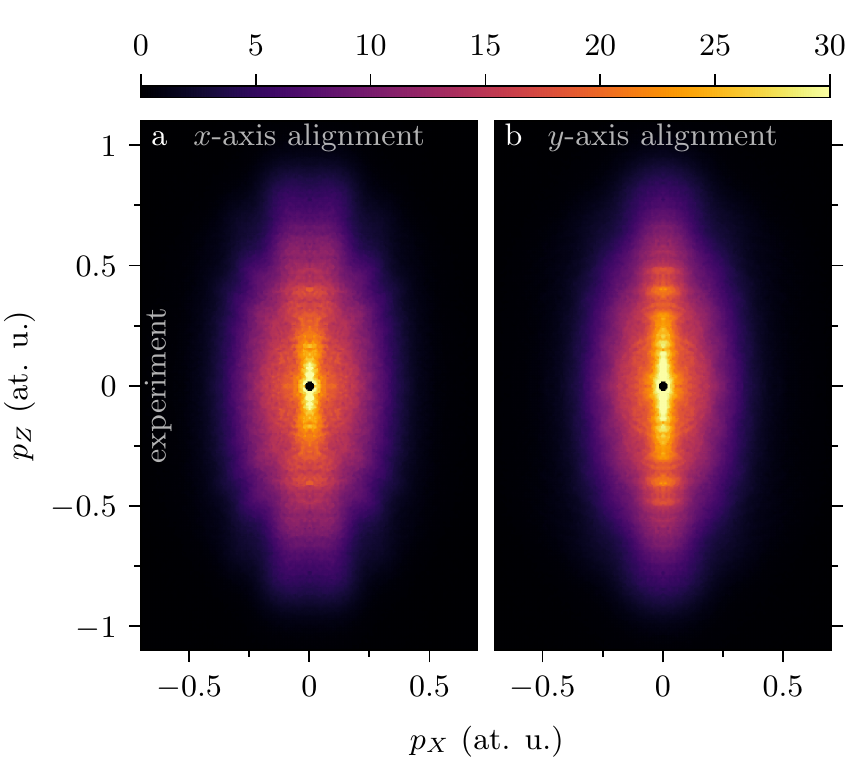}
   \caption{Experimentally obtained photoelectron-momentum maps for a) the $x$-axis alignment case
   and b) the $y$-axis alignment case. The images show a closeup of \autoref[e,~f]{fig:pmaps}
   displayed on a common linear color scale.}
   \label{fig:pmaps_zoom}
\end{figure}
\autoref{fig:pmaps} shows the calculated and measured asymptotic momentum maps. Each map exhibits an
intense part in the center ($\text{intensity}\geq0.4$, yellow to orange color), that overlaps with
two relatively faint circular patterns (intensity~$<0.4$, purple color) centered on the polarization
axis at the peak vector potential, $\abs{p_Z}=\epsilon_0/\omega\approx0.83$.

\autoref[e,~f]{fig:pmaps} show the experimentally obtained intensity-difference data sets, which
clearly differ for the two alignment scenarios employed. This difference mainly manifests itself in
the star-shaped angular structure of the intense central parts of the images. For the $x$-axis
alignment case there is an angular local minimum along the polarization axis, whereas the momentum
map for $y$-axis alignment exhibits a maximum. Generally, the angular modulations are a lot less
pronounced for $y$-axis alignment. Furthermore, both momentum maps show a few sharp radial maxima
along the $Z$~axis at momenta around $\pm0.4$, which are easier to see in the linear-scale
representation in \autoref{fig:pmaps_zoom}. The projected out-of-plane angles,
\begin{equation}
   \Omega = \arctan\left( \frac{\expectation{|p_X|}}{\expectation{|p_Z|}} \right) \,,
\end{equation}
are $\degree{25}$ for $x$- and $\degree{24}$ for $y$-axis alignment.

The simulated asymptotic momentum distributions following irradiation through a laser pulse with a
distinct temporal peak intensity are depicted in \autoref[a,~b]{fig:pmaps}. The momentum maps for
both alignment cases show rich radial and angular structure in the intense central parts. The radial
\ac{ATI} interference patterns are clearly dominant and are overall shifted in momentum between the
two cases. In the high-intensity central parts of the images a star-like pattern is visible with an
hourglass shape imprinted on it. Here, the projected out-of-plane angles for $x$- and $y$-axis
alignment are $\degree{28}$ and $\degree{26}$.

\autoref[c,~d]{fig:pmaps} show the modeled asymptotic momentum maps with focal-volume averaging,
assuming a range of the ponderomotive potential of $0.83~\hbar\omega$. Here, the \ac{ATI}
interferences are almost fully extinguished~\cite{Wiese:NJP21:083011}, which results in the
star-like angular pattern becoming the dominant feature. The angles $\Omega$ are the same as those
for the momentum maps at fixed laser intensity.

These phenomena in the momentum maps are linked to their respective individual sources within the
complete molecular strong-field ionization process: The inner, intense part of the asymptotic
momentum distribution, which constitutes the majority of the total probability density, represents
the fraction of the photoelectron wavefunction that leaves the molecule directly, \ie, without close
interaction with the cationic potential. Its continuum motion is dominated by the electric field of
the laser, giving rise to ATI interferences between adjacent optical cycles. However, the two rather
faint, disc-like shapes in the momentum maps result from strong Coulomb interaction of a small
fraction of the photoelectron wavefunction with the ionic potential upon return to the
cation~\cite{Corkum:PRL71:1994}. The outermost rings of these discs correspond to electron--cation
recollisions at maximum momentum and carry information on the molecular
geometry~\cite{Meckel:Science320:1478}. Where the inner ellipse and the outer discs overlap in
momentum space, holographic interferences can occur, which give rise to the star-like pattern
observed here. This holographic fingerprint encodes both, the tunnel exit position and the initial
momentum distribution of the photoelectron wavefunction~\cite{Huismans:Science331:61}. The embedded
hourglass, which is especially pronounced in the $x$-axis alignment case, results from the nodal
plane imprinting onto the momentum distribution at birth and thus onto the holographic pattern.

Both the out-of-plane angle $\Omega$ and the absolute kinetic energies of the \ac{ATI} interferences
provide a sensitive probe of the alignment-dependent ionization potential. While $\Omega$ is, for
identical laser parameters, mainly determined by the initial transverse momentum distribution in
\eqref{eq:psi0}, the energy of an \ac{ATI} ring,
$N\hbar\omega-\Ip{}-\Up{}$~\cite{Freeman:JPB24:325}, represents a direct link to the in-field
ionization threshold for a given number $N$ of absorbed photons.

Due to their invariance towards laser-intensity averaging the sharp maxima in the radial
distributions of the experimental momentum maps are ascribed to Freeman resonances,
resonance-enhanced multiphoton ionization through Rydberg states~\cite{Freeman:PRL59:1092,
Freeman:JPB24:325, OBrian:PRA49:R649, Wiese:NJP21:083011}. Conceptually, these electronic
resonances cannot be grasped by our current semiclassical model.

The predictive strength of the present model, in comparison to the experimental data is very good.
This is seen, for instance, by the agreement of the momentum maps from experiment and the
semiclassical model incorporating the emulation of focal-volume averaging. All features encountered
in the experimental data are qualitatively reproduced with comparable relative intensities.
Furthermore, the trend in the out-of-plane angle is reflected by the simulations, predicting a
larger angle for the $x$-axis alignment case, which can be ascribed to a stronger depression of the
ionization potential.

The hourglass shape in the $x$-axis alignment case represents the only structural deviation in a
modeled momentum map with respect to its experimental counterpart: From the semiclassical simulation
a local maximum in the angular distribution along the polarization axis is predicted, while in
experiment a local minimum is observed. This discrepancy is attributed to the nonperfect linear
polarization of the experimentally used laser pulses: A decreasing degree of linearity lowers the
return probability to the cation and thus the peculiarity of the holographic pattern, which carries
a maximum along the polarization axis. However, the imprint of the \acs{HOMO}'s nodal plane, giving
rise to a minimum along the polarization axis, remains at least as pronounced. In the limit of
circular polarization the node will be maximally distinct~\cite{Holmegaard:NatPhys6:428,
   Dimitrovski:JPB48:245601}, while the holographic pattern will be fully suppressed. Although the
description of initial in-polarization-plane momenta by means of the adiabatic tunneling theory,
including the first-order nonadiabatic correction, was established for elliptical polarization
shapes close to circular~\cite{Liu:PRL122:053202}, so far there is no equivalent theoretical
framework that tackles just slightly elliptical polarization shapes. In future experimental studies
special attention should be paid to optimizing the linearity of the laser field. This would
facilitate the comparison with simulation results and moreover maximize the re-collision probability
with the cation, which would increase the quality of holographic structures and imprints from
laser-induced electron diffraction.

\section{Conclusions}
\label{sec:conclusions}
We unraveled the strong-field photoelectron imaging of the prototypical biomolecule indole using a
combined experimental and computational approach. Strongly controlled molecules and an experimental
technique suppressing laser-intensity-volume averaging enabled the recording of
photoelectron-momentum distributions directly in the molecular frame and for a well-defined, narrow
spectrum of incident intensities. As a numerical counterpart we developed a novel, highly efficient
semiclassical model that builds upon the adiabatic tunneling theory. Both procedures revealed
holographic structures in the asymptotic momentum distributions that were found to sensitively
depend on the direction of the ionizing field's polarization axis in the molecular frame.

Based on the very good agreement between experiment and theory we are confident to have identified
all essential molecular properties that shape the photoelectron wavepacket as it is born at the
tunnel exit. Owing to the quantum-chemically exact treatment of the cation during the subsequent
continuum dynamics our model is ideally suited for studies of highly complex molecular structures
through strong-field ionization and laser-induced electron diffraction. It allows to describe
electron-diffractive imaging of biomolecules on femtosecond time scales while offering the
opportunity to fully follow the photoelectron's motion along semiclassical trajectories. Our model
description takes account of the cation's laser-induced polarization and gives rise to a
significantly faster convergence of the asymptotic photoelectron wavefunction than previously
described models~\cite{Shvetsov-Shilovski:PRA94:013415} due to the use of a spherical energy
coordinate system. Furthermore, we sketched a clear path that leads to the simplified emulation of
focal-volume averaging, enabling a direct comparison with common experimental data.

The applicability of our model is only limited by the feasibility of the quantum-chemical
computation of the molecular ion and the validity of the single-active electron approximation.
However, the simultaneous ionization from multiple orbitals could be added easily and would come
without any additional numerical expense.

Furthermore, our model could be used to simulate experiments involving circular polarization, for
example measurements of strong-field photoelectron circular dichroism, employing the corresponding
framework of the adiabatic tunneling theory with the first-order nonadiabatic
correction~\cite{Liu:PRL122:053202}.

The scattering of electrons at the quantum-chemically exact potential with the Kepler-law
approximation would also be useful for trajectory-based descriptions of conventional electron
diffraction off molecules, especially in the low-energy-electron diffraction (LEED)
regime~\cite{Pendry:LEED}.

\section*{Acknowledgments}
We thank Terry Mullins, Jean-Fran\c{c}ois Olivieri, and Andrea Trabattoni for assistance in the
experiments.

This work has been supported by the Clusters of Excellence ``Center for Ultrafast Imaging'' (CUI,
EXC~1074, ID~194651731) and ``Advanced Imaging of Matter'' (AIM, EXC~2056, ID~390715994) of the
Deutsche Forschungsgemeinschaft (DFG) and by the European Research Council under the European
Union's Seventh Framework Program (FP7/2007-2013) through the Consolidator Grant COMOTION
(ERC-Küpper-614507). J.O. gratefully acknowledges a fellowship by the Alexander von Humboldt
Foundation.

\appendix
\section{Quantum chemistry results}
\label{sec:qcres}
The equilibrium geometry of the neutral indole molecule in its electronic ground state, the
field-free ionization potential, and the permanent dipole moments and polarizability tensors of the
neutral and singly charged indole species were computed using
\texttt{GAMESS}~\cite{Schmidt:JCC14:1347, Gordon:GAMESS:2005} at the MP2/aug-cc-pVTZ level of
theory. For the sake of comparability all molecular-frame dependent quantities in this section are
given in the principal-axes-of-inertia frame, marked with a prime. However, all computations were
performed in the principal-axes-of-polarizability frame of the neutral species, which relates to the
inertial coordinate system by a rotation of $\ordsim\degree{1}$ within the molecular plane.
\begin{table}[b]
   \setlength{\tabcolsep}{5mm}
   \centering
   \begin{tabular}{rrr}
     \hline
     atom & $x'\,(\bohr{\!})$ & $y'\,(\bohr{\!})$\\
     \hline
     H & $-1.63636368$ & $-4.70545460$\\
     H &  $3.58872958$ & $-3.72370591$\\
     C & $-1.70598994$ & $-2.67828958$\\
     N &  $2.97003095$ & $-1.96352911$\\
     H & $-5.69874287$ & $-2.47547107$\\
     C &  $0.48927801$ & $-1.23389474$\\
     C & $-3.96874576$ & $-1.41779943$\\
     H &  $6.47657970$ &  $0.01897324$\\
     C &  $4.46491778$ &  $0.14961937$\\
     C &  $0.43239953$ &  $1.40456632$\\
     C & $-4.06841347$ &  $1.22748802$\\
     C &  $3.02186149$ &  $2.24100201$\\
     H & $-5.87322058$ &  $2.15064840$\\
     C & $-1.89699453$ &  $2.63858746$\\
     H &  $3.69143191$ &  $4.14478991$\\
     H & $-1.98733273$ &  $4.66494838$\\
     \hline
   \end{tabular}
   \caption{Calculated equilibrium geometry of the neutral indole molecule in its electronic ground
      state. The coordinates are given in the principal-axes-of-inertia frame. All atoms lie in the
      $z'=z=0$ plane. The same atomic coordinates were used for the description of the indole cation
      during the semiclassical computations.}
   \label{tab:eqgeo}
\end{table}
\autoref{tab:eqgeo} lists the atomic coordinates in the resulting equilibrium geometry, which is in
excellent agreement with previously performed calculations at CC2/cc-pVTZ level of
theory~\cite{Brand:PCCP12:4968:2010}. The field-free ionization potential was calculated as the
difference of the field-free total energies of cation and neutral molecule
$\Ip{}^{(0)}=U_c^{(0)}-U_n^{(0)}=8.29~\text{eV}$ and is in reasonable agreement with the
experimentally measured vertical ionization energy of 7.9~eV~\cite{NIST:webbook:2017}. The permanent
dipole moments of the neutral and cation ground states are
\begin{align*}
  \vec{\mu}'_n
  & = \begin{pmatrix}
     0.5804\\
     -0.5776\\
     0\\
  \end{pmatrix}~\text{at. u.}
  = \begin{pmatrix}
     1.475\\
     -1.468\\
     0\\
  \end{pmatrix}~\text{D} \\
  \vec{\mu}'_c
  &= \begin{pmatrix}
     0.5043\\
     -0.5838\\
     0\\
  \end{pmatrix}~\text{at. u.}
  = \begin{pmatrix}
     1.282\\
     -1.484\\
     0\\
  \end{pmatrix}~\text{D} \,.
\end{align*}
$\vec{\mu}'_n$ is in excellent agreement with experimental
observations~\cite{Caminati:JMolStruct240:253, Kang:JCP122:174301} and quantum-chemistry
calculations performed at the CC2/cc-pVTZ level of theory~\cite{Brand:PCCP12:4968:2010}. The
corresponding static polarizability tensors are \nopagebreak
\begin{align*}
  \alpha'_n
  & = \begin{pmatrix}
     135.2 & 1.678 & 0\\
     1.678 & 104.0 & 0\\
     0     & 0     & 59.43\\
  \end{pmatrix}~\text{at. u.} \\
  & = \begin{pmatrix}
     20.04 & 0.2487 & 0\\
     0.2487 & 15.41 & 0\\
     0     & 0     & 8.808\\
  \end{pmatrix}~10^{-3}~\text{nm}^3 \\
  \alpha'_c & = \begin{pmatrix}
     240.9 & 38.56 & 0\\
     38.56 & 105.3 & 0\\
     0     & 0     & 49.55\\
  \end{pmatrix}~\text{at. u.} \\
  & = \begin{pmatrix}
     35.70 & 5.717 & 0\\
     5.717 & 15.61 & 0\\
     0     & 0     & 7.344\\
  \end{pmatrix}~10^{-3}~\text{nm}^3 \,.
\end{align*}
The isotropic polarizability of the neutral species,
$\tr(\alpha'_n)/3=14.75\cdot10^{-3}~\text{nm}^3$, is in good agreement with previous quantum
chemistry results~\cite{NIST:CCCBDB}.

\section{Field-direction model}
\label{sec:fielddir}
\begin{figure}
   \includegraphics[width=\linewidth]{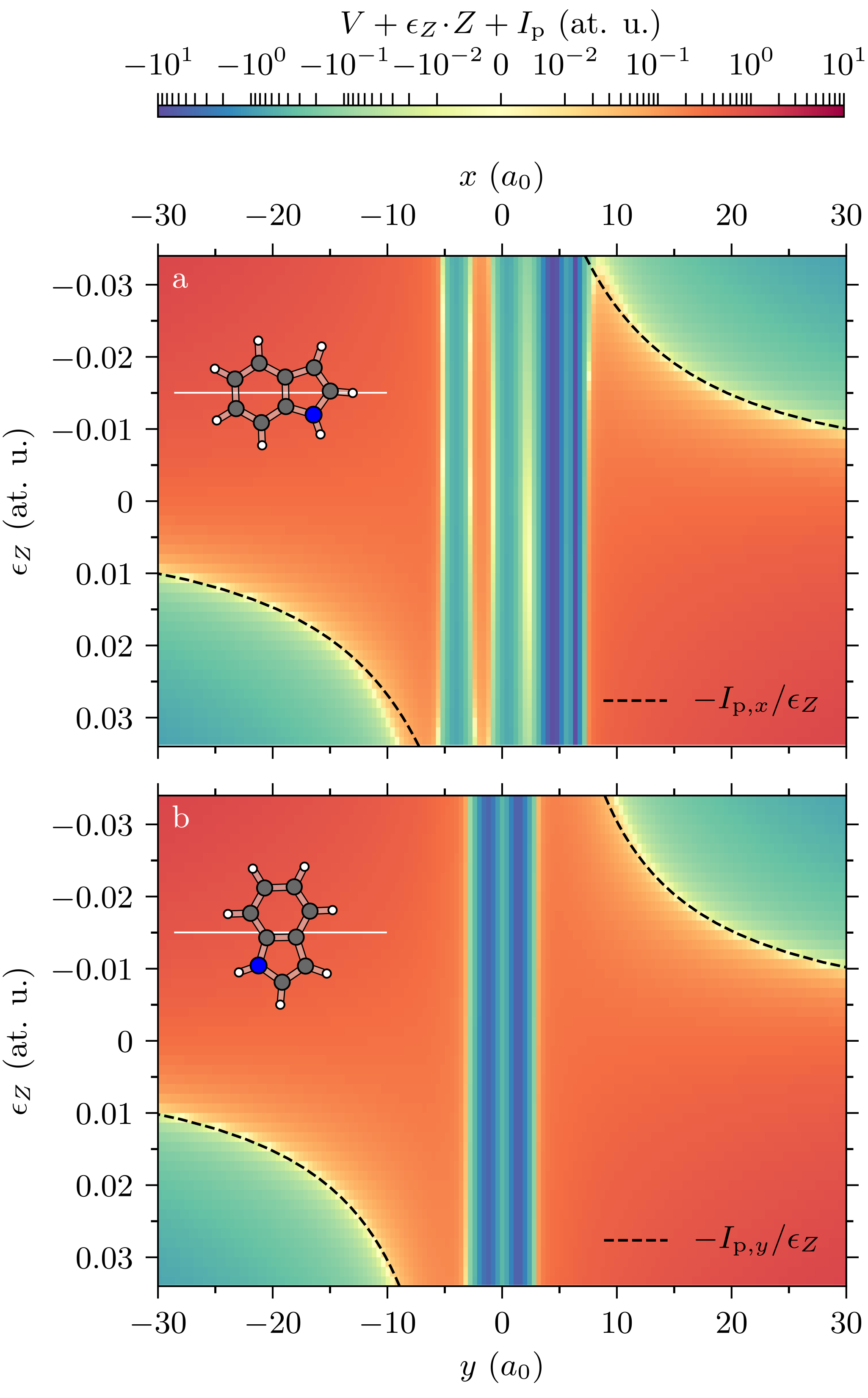}
   \caption{The sum of quasistatic molecular potential, laser potential, and ionization energy is
      shown for a) $x$-axis and b) $y$-axis alignment; each panel depicts a single orientation. Only
      positions on the one principal polarizability axis were evaluated that coincided with the
      polarization axis, which is depicted as white line within the corresponding molecule sketch.
      The deep potential wells near the center of nuclear charge reflect the proximity to the
      closest nuclei. For comparison, the respective tunnel exit positions for a quasistatic
      laser-electric field and neglecting the Coulomb distortion of the
      barrier~\cite{Popruzhenko:JPB47:204001} are shown as dashed black lines.}
   \label{fig:fielddir}
\end{figure}
The tunnel-exit positions were obtained in the field-direction model at the HF/aug-cc-pVTZ level of
theory using \texttt{Psi4}~\cite{Parrish:JCTC13:3185}. The quasistatic electric potential of the
neutral indole molecule, $V(\vec{\epsilon})$, was computed along the alignment-dependent molecular
axis that coincided with the polarization axis of the ionizing laser field. That is, only positions
on the $x$~axis were evaluated for the $x$-axis alignment case and only positions on the $y$~axis
for $y$-axis alignment. In both cases the sampling line was chosen to cross the molecule's center of
nuclear charge. Position and external electric field were sampled in intervals of
$\Delta{r}=\bohr{0.5}$ and $\Delta\epsilon=1.4\times10^{-3}$. Subsequently, the sum of quasistatic
potential, laser potential, and field-dependent ionization energy,
$V(\vec{\epsilon})+\epsilon_Z\cdot{Z}+\Ip{}(\vec{\epsilon})$, was examined, which assumes a value of
$0$ at the tunnel exit. \autoref{fig:fielddir} illustrates the corresponding results for the two
alignment scenarios. The actual tunnel-exit position used in the semiclassical computations was
eventually determined by finding the root of
$V(\vec{\epsilon})+\epsilon_Z\cdot Z+\Ip{}(\vec{\epsilon})$ at the given instantaneous
laser-electric field, which appears as a yellow seam in \autoref{fig:fielddir}. For the $x$-axis
alignment case and $\epsilon_Z\approx-3\times10^{-2}$ the maximum of the potential barrier was found
to be suppressed below 0, \ie, in the barrier-suppression regime. For these cases the barrier
maximum was used as tunnel exit position and the energy difference to 0 was assumed to be
transformed into additional longitudinal momentum opposing the direction of the instantaneous
laser-electric field.

\section{Cationic electric field}
\label{sec:catfield}
The external-field dependent cationic field of the indole molecule,
$-\vec{\nabla}V(\vec{\epsilon},\vec{r})$, was calculated at the HF/aug-cc-pVTZ level of theory using
\texttt{Psi4}~\cite{Parrish:JCTC13:3185}. Positions were sampled in intervals of
$\Delta{r}=\bohr{0.2}$ and the laser-electric field was sampled in steps of
$\Delta\epsilon=2.8\times10^{-3}$ over a $(\bohr{60})^3$ cube centered at the cation's center of
nuclear charge.

\bibliography{string,cmi}
\end{document}